\documentclass[12pt]{iopart}
\usepackage{iopams}

\newcommand{\no}{\nonumber\\}
\newcommand{\be}{\begin{equation}}
\newcommand{\ee}{\end{equation}}
\newcommand{\ba}{\begin{eqnarray}}
\newcommand{\ea}{\end{eqnarray}}

\newcommand{\la}[1]{\label{#1}}

\def\gl#1{(\ref{#1})}

\date{}
\begin{document}
\title[Non-Hermitian Quantum Mechanics]{Non-Hermitian Quantum Mechanics of Non-diagonalizable
Hamiltonians: puzzles with self-orthogonal states}
\author{A V Sokolov\dag\ , A A Andrianov\dag\ and F Cannata\ddag}
\address{\dag\ V A Fock Institute of Physics, Sankt-Petersburg
State University,
198504 Sankt-Petersburg, Russia}
\address{\ddag\
Dipartmento di Fisica and INFN, Via Irnerio 46, 40126 Bologna,
Italy}

\ead{sokolov@pdmi.ras.ru;\quad andrianov@bo.infn.it;\quad cannata@bo.infn.it}

\begin{abstract} We consider QM with non-Hermitian quasi-diagonalizable
Hamiltonians, {\it i.e.} the  Hamiltonians
having a number of Jordan cells in particular biorthogonal bases. The "self-orthogonality" phenomenon is clarified in terms of a correct spectral decomposition  and
it is shown  that "self-orthogonal" states never jeopardize resolution of identity and thereby quantum averages of observables. The example of a complex potential leading to one Jordan cell in the Hamiltonian  is constructed and its origin from level coalescence is elucidated.
Some puzzles with zero-binorm bound states in continuous
spectrum are unraveled with the help of a correct resolution of identity.
\end{abstract}

\submitto{Journal of Physics A: Math. Gen.}
\pacs{03.65.-w,03.65.Ca,03.65.Ge}

\maketitle

\section{Introduction}
The variety of complex potentials in Quantum Physics is associated
typically with  open systems when a control of information
is partially lost and thereby the unitarity of observable evolution is broken.  For this class of quantum systems
the energy eigenvalues may have an
 imaginary part which signals the opening of new channels not directly measured in a given experiment. In this context non-Hermitian interactions have been used in Field Theory and
Statistical Mechanics for many years with
applications to Condensed Matter, Quantum Optics and Hadronic and Nuclear Physics \cite{optical} -- \cite{muga}.  The subject of non-self-adjoint operators
has been also under intensive mathematical investigations \cite{dunf,pavl,davies}, in particular, interesting examples of non-Hermitian effective Hamiltonian operators have been found for the Faddeev equations \cite{yak}.

An important
 class of complex Hamiltonians deals with a real spectrum \cite{complex0,acdicom}, in particular, in the PT-symmetric Quantum Mechanics \cite{bender}--\cite{ventura}
and its pseudo-Hermitian generalization \cite{mostaf,mostaf1}. Scattering problems for such Hamiltonians have been investigated in
\cite{muga,levai} .

For complex, non-Hermitian potentials the natural spectral decomposition  exploits the sets of biorthogonal states \cite{curt} ,  and within this framework one can discover new features that never happen for closed systems with  Hermitian Hamiltonians
possessing real spectrum\footnote{An exception concerns the action of the Hamiltonian operator on  zero-mode subspaces of supercharges in Nonlinear SUSY \cite{ais} -- \cite{fern}. For confluent NSUSY, a  Hermitian Hamiltonian may produce a  non-Hermitian matrix, with Jordan cells \cite{ansok,ancan} after quasi-diagonalization. In this case zero-mode subspaces of supercharges include also non-normalizable
solutions of the Schr\"odinger equation which don't belong to the energy spectrum of the original self-adjoint Hamiltonian.} :
namely, certain Hamiltonians may not be diagonalizable \cite{solov} with a help of biorthogonal bases  and can be reduced only to a quasi-diagonal form with a number of Jordan cells \cite{mostaf}.  Such a feature can be realized by level crossing which, in fact, occurs (after some kind of complexification)  in atomic and molecular spectra \cite{solov} and Optics \cite{berry} (see more examples in  \cite{mois2}) as well as in PT-symmetric quantum systems \cite{znojil,dorey,sams1}.
In this case some eigenstates seem to be "self-orthogonal" in respect to a binorm \cite{mois2,mois1} .  The latter quite intriguing phenomenon has been interpreted as
a sort of phase transition \cite{mois2} .

The main purpose of the present work is to clarify the "self-orthogonality" in terms of a correct spectral decomposition both for discrete and for continuous spectra and to show that, at least, in one-dimensional Quantum Mechanics such
states never jeopardize  resolution of identity for the discrete or bound state spectrum  and thereby don't affect quantum averages of observables.

We start introducing the notion of biorthogonal basis and, correspondingly, the resolution of identity for a non-Hermitian diagonalizable Hamiltonian. In Sec. 2 the appearance of associated functions is discussed and in Sec. 3 non-diagonalizable (but quasi-diagonalizable) Hamiltonians with finite-size Jordan cells are analyzed.  Special attention is paid to the definition of a biorthogonal {\it diagonal} basis  and the meaning of zero-binorm states is clarified. Namely, it is shown that the apparent self-orthogonality of eigenfunctions and associated functions is misleading as they never replicate themselves  as relative pairs in diagonal resolution of identity. Instead, the "self-orthogonality" involves the different elements in the related basis thereby being addressed to a conventional orthogonality. The construction of such biorthogonal bases with pairs of mutually complex-conjugated base functions is described .
In Sec. 4 another representation of non-diagonalizable Hamiltonians , manifestly
symmetric under transposition is given, compatible with a diagonal resolution of identity. In Sec. 5
the example of a (transparent) complex potential leading to the non-diagonalizable Hamiltonian
 with one Jordan $2\times2$ cell is constructed and its origin from level coalescence is illustrated.

 On the other hand some puzzles with zero-binorm bound states arise in continuous
spectrum and they  are unraveled in Sec.6 with the help of a correct resolution of identity. Its proof is relegated to the Appendix.
 In Sec. 7 we complete our analysis with discussion of singularities in the spectral parameter for
resolvents and of scattering characteristics for previous examples.
We conclude with some proposals for probabilistic interpretation of wave functions defined in respect to a biorthogonal basis which does not allow any
negative or zero-norm states. 

There are certain links of our approach to  the 
works \cite{gadella} on  Jordan cells associated with the occurrence of non-Hermitian degeneracies for essentially Hermitian Hamiltonians where the description has been developed  for complex eigenvalue Gamow states (resonances) unbounded in their asymptotics and, in general, not belonging to the Hilbert space. On the contrary, we instead examine Nonhermitian Hamiltonians with normalizable bound  and associated states.

In our paper we deal with complex one-dimensional potentials $V(x)
\not= V^*(x)$ and
respectively with non-Hermitian Hamiltonians $h$ of Schr\"odinger type,
defined on the real axis,
\be h \equiv - \partial^2_x + V(x) ,\ee
which are assumed to be $t$-symmetric or self-transposed under
the $^t$ -- transposition operation, $h = h^t $.
Only scalar local potentials  will be analyzed which are
obviously symmetric under transposition (for some matrix non-diagonalizable problems, see \cite{mois1,matr}).
Throughout this work the units will be used  with $m=1/2,\ \hbar = 1, c=1$ which leads to dimensionless energies .

Let us first define a class of one-dimensional non-Hermitian  {\it diagonalizable}
Hamiltonian $h$ with discrete spectrum  such  that:\\
a)  a biorthogonal system $\{|\psi_n\rangle, |\tilde\psi_n\rangle\}$ exists,
\be
\fl h|\psi_n\rangle = \lambda_n|\psi_n\rangle,\qquad h^\dag|\tilde\psi_n\rangle=
\lambda_n^*|\tilde\psi_n\rangle,\qquad\langle\tilde\psi_n|\psi_m\rangle=
\langle\psi_m|\tilde\psi_n\rangle=\delta_{nm},\ee
b) the complete resolution of identity in terms of these  bases and the spectral decomposition of the Hamiltonian
 hold,
\be I=\sum\limits_n|\psi_n\rangle\langle\tilde\psi_n|,\qquad h=
\sum\limits_n\lambda_n|\psi_n\rangle\langle\tilde\psi_n|.\ee

In the coordinate representation,
\be \psi_n(x)=\langle x|\psi_n\rangle,\qquad\tilde\psi_n(x)=\langle x|
\tilde\psi_n\rangle ,\ee
the resolution of identity has the form,
\be \delta(x-x')=\langle x'|x\rangle=\sum\limits_n\psi_n(x')\tilde\psi_n^*(x).\ee
The differential equations,
\be h\psi_n=\lambda_n\psi_n, \qquad h^\dag\tilde\psi_n=\lambda_n^*\tilde\psi_n , \ee
and the fact that there is only one normalizable eigenfunction of $h$ for the
eigenvalue $\lambda_n$ (up to a constant factor), allow one to conclude that
\be \tilde\psi_n^*(x)\equiv\alpha_n\psi_n(x),\qquad\alpha_n={\rm{Const}}\ne0.\ee
Hence the  system  $\{|\psi_n\rangle, |\tilde\psi_n\rangle\}$ can be
redefined
\be |\psi_n\rangle\to{1\over\sqrt{\alpha_n}}|\psi_n\rangle,\qquad
|\tilde\psi_n\rangle\to\sqrt{\alpha_n^*}|\tilde\psi_n\rangle , \label{normal} \ee
so that
\be\tilde\psi_n^*(x)\equiv\psi_n(x),\qquad\int\limits_{-\infty}^{+\infty}\psi_n(x)
\psi_m(x)\,dx=\delta_{nm}. \la{bior}\ee
We stress that the non-vanishing binorms in Eq.\gl{bior} support the completeness of this basis,
{\it i.e.} the resolution of identity,
\be
\qquad\delta(x-x')=\sum\limits_n
\psi_n(x)\psi_n(x') . \label{decom}
\ee
Indeed if some of the states in Eq. \gl{decom} were "self-orthogonal" (as it has been accepted in
\cite{mois1}) , {\it i.e.} had zero binorms in \gl{bior}, the would-be unity in \gl{decom} would annihilate such states
thereby signalling the incompleteness.
\section{Non-diagonalizable Hamiltonians and zero-\-binorm states}
For complex Hamiltonians one can formulate the extended eigenvalue problem,
searching not only for normalizable eigenfunctions but also for
normalizable associated functions for discrete part of the energy
spectrum. Some related
problems have
been known for a long time in mathematics of linear differential
equations
 (see for
instance, \cite{naim}) .

Let us give  the formal definition.\\

{\bf Definition.} \quad The function $\psi_{n,i}(x)$ is called a formal
associated
function of $i$-th order of the Hamiltonian $h$ for a spectral value
$\lambda_n$, if
\be (h-\lambda_n)^{i+1}\psi_{n,i}\equiv0,\qquad
(h-\lambda_n)^{i}\psi_{n,i}\not\equiv 0, \label{canbas1}\ee
where 'formal' emphasizes that a related function is not necessarily
normalizable.
\bigskip

In particular, the associated function of zero order $\psi_{n,0}$ is
a formal eigenfunction of $h$ (a solution of the homogeneous
Schr\"odinger equation, not necessarily normalizable).

Let us single out normalizable associated functions and the case
when $h$ maps them into  normalizable functions. Evidently this
may occur only for non-Hermitian Hamiltonians. Then for any
normalizable associated functions $\psi_{n,i}(x)$ and $\psi_{n',i'}
(x)$ the transposition symmetry holds
\be\int\limits_{-\infty}^{+\infty}h \psi_{n,i}(x)
\psi_{n',i'}(x)\,dx=\int\limits_{-\infty}^{+\infty} \psi_{n,i}(x) h
\psi_{n',i'}(x)\,dx . \ee Furthermore one can prove the following
relations: \be \int\limits_{-\infty}^{+\infty} \psi_{n,i}(x)
\psi_{n',i'}(x) \,dx\equiv(\psi^*_{n,i} \ , \psi_{n',i'}) =0,
\qquad\lambda_n\ne\lambda_{n'},\ee
where $(\ldots,\ldots)$ is scalar product.

As well, let's take two normalizable associated functions $\psi_{n,k}(x)$ and
$\psi_{n,k'}(x)$ so that, in general, $k\not= k' $ and there are two different sequences of associated functions
for $i\leq k$ and $i'\leq k'$
\be \psi_{n,i}(x)= (h-\lambda_n)^{k-i}
\psi_{n,k}(x), \quad \psi_{n,i'}(x)= (h-\lambda_n)^{k'-i'}
\psi_{n,k'}(x) .\label{canbas2}\ee
Then
\be \fl \int\limits_{-\infty}^{+\infty}
\psi_{n,i}(x) \psi_{n,i'}(x)\,dx=(\psi^*_{n,i} \ , \psi_{n,i'}) =0 ,\qquad  i+i'\le\max\{k,k'\}-1. \label{binorm}
\ee
In particular, for some normalizable
associated function $\psi_{n,l}(x)$, the "self-orthogonality" \cite{mois1} is
realized ,
\be
\fl \int\limits_{-\infty}^{+\infty}\psi^2_{n,l}(x) \,dx=0, \qquad
\psi_{n,l}(x) =(h-\lambda)^{i-l}\psi_{n,i}(x), \qquad l=0,\ldots,
\Big[{{i-1}\over2}\Big].\ee
Thus, when assigning \cite{mois2} the probabilistic meaning for the binorm $(\Psi^*, \Psi)$ one comes to
conclusion that a sort of intriguing phase transition occurs in such a system, signalled by the  puzzling divergence of some averages of observables,
\be
\langle O\rangle = \frac{(\psi^*_{n,l}\ , O\psi_{n,l})}{(\psi^*_{n,l}\ , \psi_{n,l} )} = \frac{\int\limits_{-\infty}^{+\infty}\psi_{n,l}(x)\  O\psi_{n,l} (x) dx}{\int\limits_{-\infty}^{+\infty}\psi_{n,l}^2 (x) dx }\rightarrow \infty .
\label{aver}\ee
All the above relations are derived from the symmetry of a Hamiltonian under transposition and the very
definition of associated functions and therefore the existence of self-orthogonal states seems to be inherent for any non-diagonalizable Hamiltonians with normalizable associated functions.
\section{Towards resolution of puzzle with self-\-ortho\-go\-nal states for Hamiltonians with finite-size Jordan cells}
Let us  show that the puzzle with self-orthogonal states may appear, in fact, due to misinterpretation
of what are the pairs of orthogonal states in a true biorthogonal basis.
We proceed to  the special class of Hamiltonians  for which
 the spectrum is discrete and there is a complete
biorthogonal system
 $\{|\psi_n,a,i\rangle,|\tilde\psi_n,a,i\rangle\}$ such
that,
\ba \fl && h|\psi_n,a,0\rangle=\lambda_n|\psi_n,a,0\rangle, \qquad(h-\lambda_n)|\psi_n,a,i\rangle
=|\psi_n,a,i-1\rangle , \label{schr1}\\
\fl && h^\dagger|\tilde\psi_n,a,p_ {n,a}-1\rangle=\lambda^*_n|\tilde\psi_n,a,p_{n,a}-1\rangle,
\quad(h^\dagger-\lambda^*_n)|\tilde\psi_n,a,p_{n,a}- i -1\rangle
=|\tilde\psi_n,a,p_{n,a} -i\rangle, \nonumber \ea
where
$n=0, 1, 2, \ldots$ is an index of an $h$ eigenvalue $\lambda_n,$\\
$a=1$, \dots, $d_n$ is an index of a Jordan cell (block) for the given
eigenvalue,
$\lambda_n$;\\ $d_n$ is a number of Jordan cells for $\lambda_n$;\\
$i=0$, \dots, $p_{n,a}-1$ is an index of associated function in the Jordan cell
with indexes $n,a$\\
and $p_{n,a}$ is a dimension of this Jordan cell. We have taken a general framework which is applicable also for matrix and/or multidimensional Hamiltonians. But the main results of this and the next sections are guaranteed only for
scalar one-dimensional Hamiltonians with local potentials.

We remark that the number $d_n$
is called a geometric multiplicity of the eigenvalue $\lambda_n$ .
For a scalar one-dimensional Schr\"odinger equation
it cannot normally exceed 1 (but  may reach 2 in specific cases of periodic potentials and of potentials unbounded from below).  In turn,  the sum $\sum_a
p_{n,a}$ is called an algebraic multiplicity of the eigenvalue $\lambda_n$.

The completeness implies the biorthogonality relations (in line with the enumeration of states $|\tilde\psi_n,a,i\rangle$
given in Eq.\gl{schr1})
\be\langle\tilde\psi_n,a,i|\psi_m,b,j\rangle=\delta_{nm}\delta_{ab}
\delta_{ij} \ ,
\label{assorth}\ee
and the resolution of identity
\be I=\sum\limits_{n=0}^{+\infty}\sum\limits_{a=1}^{d_n}\sum\limits_{i=0
}^{p_{n,a}-1}|\psi_n,a,i\rangle\langle\tilde\psi_n,a,i| .\ee
The spectral decomposition for the Hamiltonian can be constructed as well,
\be h=\sum\limits_{n=0}^{+\infty}\sum\limits_{a=1}^{d_n}\Big[\lambda_n
\sum\limits_{i=0
}^{p_{n,a}-1}|\psi_n,a,i\rangle\langle\tilde\psi_n,a,i|+
\sum\limits_{i=0
}^{p_{n,a}-2}|\psi_n,a,i\rangle\langle\tilde\psi_n,a,i+1|\Big].\ee
It represents the analog of the block-diagonal Jordan form for
arbitrary non-Hermitian matrices \cite{gant}.

If existing such biorthogonal systems are not unique. Indeed the relations \gl{schr1}
remain invariant under the group of triangle transformations,
\ba
&& |\psi'_n,a,i\rangle = \sum\limits_{0\leq j \leq i} \alpha_{ij}|\psi_n,a,j\rangle,\nonumber\\
&&|\tilde\psi'_n,a,k\rangle =\sum\limits_{k\leq l\leq p_{n,a}-1} \beta_{kl}|\tilde\psi_n,a,l\rangle,
\label{trian}
\ea
where the matrix elements must obey the following equations,
\ba
&& \alpha_{ij}= \alpha_{i+1,\ j+1} = \alpha_{i-j,\ 0},\quad \alpha_{00}\not= 0 , \nonumber\\
&&\beta_{kl}= \beta_{k+1,\ l+1}=\beta_{k-l+p_{n,a}-1,\
p_{n,a}-1} ,\quad \beta_{p_{n,a}-1,\
p_{n,a}-1}\not= 0 . \ea
The biorthogonality \gl{assorth} restricts the choice of pairs
of matrices $\hat\alpha$ and $\hat\beta$  in \gl{trian} to be,
\be \hat\alpha
\hat\beta^\dagger = \hat\beta^\dagger \hat\alpha =\mathbb{I}.
\ee
This freedom in the redefinition of the
biorthogonal basis is similar to Eq. \gl{normal} and it can be exploited to
define the pairs of biorthogonal functions $\psi_{n,a,i} (x) \equiv \langle x |\psi_n,a,i\rangle$ and
$\tilde\psi_{n,a,i} (x) \equiv \langle x|\tilde\psi_n,a,i\rangle$ in accordance with \gl{bior}.
 However
one has to take into account our enumeration of associated functions
$\psi_{n,a,i} (x)$ vs. their conjugated ones
$\tilde\psi_{n,a,i} (x)$ as it is
introduced in Eqs. \gl{schr1}
\be \psi_{n,a,i} (x) =
\tilde\psi^*_{n,a,p_{n,a}-i-1} (x) \equiv \langle\tilde
\psi_n,a,p_{n,a}-i-1|x\rangle . \label{special} \ee
Then the analog of  Eq. \gl{bior} reads,
\be
\int\limits_{-\infty}^{+\infty} \psi_{n,a,i} (x) \psi_{m,b,p_{m,b}-j-1} (x) dx = \delta_{nm}\delta_{ab}
\delta_{ij}\  .
\ee
We stress that this kind of
biorthogonal systems is determined uniquely  up to an overall sign.

In these terms it becomes clear that the relations \gl{binorm} have
the meaning of orthogonality of some off-diagonal pairs in the biorthogonal
system $\{|\psi_n,a,i\rangle,|\tilde\psi_n,a,j\rangle\}$ as
$$
\psi_{n,a,i} (x) = (h - \lambda_n)^{p_{n,a}-1 -i} \psi_{n,a, p_{n,a}-1} (x),$$
\be \tilde\psi^*_{n,a,j} (x)  = \psi_{n,a, p_{n,a}-1-j} (x) =
(h - \lambda_n)^{j} \psi_{n,a, p_{n,a}-1} (x). \label{chain}
\ee
When comparing with specification of indices in Eq. \gl{binorm} one identifies\\ $p_{n,a}-1-j \leftrightarrow i,\quad i \leftrightarrow i' $.
In both cases $k= k' = p_{n,a}-1$ . Then the inequality \gl{binorm} singles out
 off-diagonal binorms, $i \leq j-1$.  From Eq. \gl{chain} it follows 
that in order to have all diagonal binorms non-vanishing it is sufficient
to prove  that at least one of them is not zero because
\ba
&&\int\limits_{-\infty}^{+\infty} \psi_{n,a,0}(x)
\psi_{n,a, p_{n,a}-1}(x)\,dx \no &&= \int\limits_{-\infty}^{+\infty} \Big[(h - \lambda_n)^{p_{n,a}-1} \psi_{n,a, p_{n,a}-1} (x)\Big]
\psi_{n,a, p_{n,a}-1}(x)\,dx \no &&= \int\limits_{-\infty}^{+\infty} \psi_{n,a,i}(x)
\psi_{n,a, p_{n,a}-1-i}(x)\,dx\not= 0 .
\ea
The latter is necessary for the completeness of the basis ( because of the absence of self-orthogonal pairs of basis elements made of bound and associated functions when resolution of identity is diagonal).

Going back to the definition of quantum-state averages of certain observables  we realize that
the matrix element used in \gl{aver} is {\it not diagonal} and therefore this relation cannot be interpreted
as an average (compare with \cite{mois2,mois1}) of a putative order-parameter
like operator.
\section{$t$-symmetric representation of non-diagonalizable Hamiltonians}

We still notice that the biorthogonal basis \gl{special} does not provide a manifestly $t$-symmetric representation of the Hamiltonian (which is  symmetric  under transposition $h = h^t$ in the coordinate representation as a finite-order differential operator).
One can obtain another biorthogonal basis using the canonical set  of (normalizable) associated functions given by Eq. \gl{schr1} and their complex conjugates in an analogy to \gl{normal} . It can be achieved  by means of renumbering of conjugated elements of the biorthogonal
system \gl{schr1} ,
\be |\hat\psi_n,a,j\rangle=|\tilde\psi_n,a,p_{n,a}-j-1\rangle,\qquad
j=0,\dots,p_{n,a}-1 .\ee
Eventually one arrives to the $t$-{\it symmetric} spectral decomposition for $h$:
\be \fl h=\sum\limits_{n=0}^{+\infty}\sum\limits_{a=1}^{d_n}\Big[\lambda_n
\sum\limits_{j=0
}^{p_{n,a}-1}|\psi_n,a,j\rangle\langle\hat\psi_n,a,p_{n,a}-j-1|+
\sum\limits_{i=0
}^{p_{n,a}-2}|\psi_n,a,j\rangle\langle\hat\psi_n,a,p_{n,a}-j-2|\Big] ,\ee
which looks like a Jordan decomposition along the secondary diagonal. Evidently
in the coordinate representation
the Hamiltonian operator is manifestly $t$-symmetric when the special biorthogonal basis \gl{special},
\be \psi_{n,a,j} (x) =
\hat\psi^*_{n,a,j} (x) \equiv \langle\hat
\psi_n,a,j|x\rangle , \label{specialhat} \ee
 is chosen.

But the resolution of identity in this case is not diagonal,
\be I=\sum\limits_{n=0}^{+\infty}\sum\limits_{a=1}^{d_n}\sum\limits_{j=0
}^{p_{n,a}-1}|\psi_n,a,j\rangle\langle\hat\psi_n,a,p_{n,a}-j-1| \ee
although $t$-symmetric. One can diagonalize this resolution of identity by a non-degenerate orthogonal transformation $\Omega$ of sub-bases
 in each non-diagonal sub-block,
\be
|\psi_n,a,j\rangle =  \sum\limits_{k=0
}^{p_{n,a}-1}\Omega_{jk}|\psi'_n,a,k\rangle ,\quad \langle\hat\psi_n,a,j| =  \sum\limits_{k=0
}^{p_{n,a}-1}\Omega_{jk}\langle\hat\psi'_n,a,k|
\ee
retaining the type of the basis \gl{specialhat} .
Then one finds a number eigenvalues $\pm 1$. In order to come to the canonical form of a basis \gl{assorth} one has to rotate by the complex unit $i$ the pairs in the basis \gl{specialhat} normalized on $-1$  . Evidently the combination of the transformation $\Omega$ and  such a rotation contains complex elements and is not orthogonal.

The remaining freedom of basis redefinition with the help of orthogonal rotations cannot
provide the consequent diagonalization of the symmetric Hamiltonian matrix in each non-diagonal block. The reason is
that some of eigenvectors of the Hamiltonian sub-matrices have  zero binorms, in particular, those ones which are related to the true Hamiltonian eigenfunctions. Thus while being a $t$-symmetric
operator with symmetric matrix representation, the Hamiltonian remains essentially non-diagonalizable \footnote{We notice that a similar symmetric representation for the Hamiltonian has been exemplified in a specific model with one eigenstate and one associated function\cite{sams1} }.

We remark that in the general
case the existence and the completeness of a  biorthogonal system  is not obvious
(especially if the continuous spectrum is present) and needs a careful examination. In particular,
at the border between discrete and continuous spectra  and in the
continuous spectrum itself one
can anticipate to have puzzling  states with non-trivial  role in the
spectral decomposition. This peculiarities will be discussed in the next Sections.

\section{A model with Non-diagonalizable Hamiltonian and its origin from level coalescence}
\def\sh{{\rm{sh}}\,}\def\ch{{\rm{ch}}\,}\def\th{{\rm{th}}\,}
In this Section we build a model \footnote{All Hamiltonians
considered
in this and the next sections can be constructed with the help of SUSY methods \cite{cooper,ansok,ancan} and
are
intertwined with the Hamiltonian of a free particle by differential
operators
of the 1st or 2nd order.} with non-Hermitian Hamiltonian which has a real continuous spectrum and, in addition,  possesses a Jordan cell spanned on the bound
state and a normalizable associated state. This model does not belong to the class of Hamiltonians with purely discrete spectrum
considered in the preceding sections but being block-diagonal it inherits some of their properties in the bound state sector.  Further on we demonstrate how this kind of degeneracy arises from
coalescence of a pair of non-degenerate levels.

\subsection{Jordan cell for bound state}
 The model Hamiltonian contains the potential  with coordinates selectively shifted into complex plane,
\be
\fl h=-\partial^2-16\alpha^2{{\alpha(x-z)\sh(
2\alpha x)-2\ch^2(\alpha x)}\over{[\sh(2\alpha x)+
2\alpha(x-z)]^2}},\qquad\alpha>0,\quad z\in \mathbb{C},
\quad {\rm{Im}}\,z\ne0 .\label{ham1}\ee
This Hamiltonian  is not PT-symmetric unless ${\rm{Re}}\, z = 0$ .
It has the Jordan cell, spanned by the normalizable eigenfunction $\psi_{0}(x)$
and
associated function $\psi_{1}(x)$ on the level $\lambda_1=-\alpha^2$,
\be \fl \psi_{0}(x)={{(2\alpha)^{3/2}\ch(\alpha x)}\over{\sh(2\alpha
x)+2\alpha(x-z)}},\qquad
\psi_{1}(x)={{{{2\alpha}{(x-z)}}\sh(\alpha x)-
\ch(\alpha x)}\over{\sqrt{2\alpha}[\sh(2\alpha x)+2\alpha(x-z)]}},\ee
\be h\psi_{0}=\lambda_1\psi_{0},\qquad(h-\lambda_1)
\psi_{1}=\psi_{0}.\ee
In turn, the eigenfunctions of $h$ for continuous spectrum read,
\be \psi(x;k)={1\over\sqrt{2\pi}}\left[1+{{ik}\over{\alpha^2+k^2}}
{{W'(x)}\over{W(x)}}-{1\over{2(\alpha^2+k^2)}}{{W''(x)}\over{W(x)}}
\right]e^{ikx},\la{psixk1}\ee
$$k\in\mathbb R, \qquad h\psi(x;k)=k^2\psi(x;k),\qquad W(x)=\sh(2\alpha
x)+2\alpha(x-z).$$
The eigenfunctions and the associated function of $h$ obey
the biorthogonality relations,
\ba &&\int\limits_{-\infty}^{+\infty}\psi_{0,1}^2(x)\,dx=0,\qquad
\int\limits_{-\infty}^{+\infty}\psi_{0}(x)\psi_{1}(x)\,dx=1,\no &&
\int\limits_{-\infty}^{+\infty}\psi_{0,1}(x)\psi(x;k)\,dx=0 ,\qquad
 \int\limits_{-\infty}^{+\infty}\psi(x;k)\psi(x;-k')\,dx=
\delta(k-k') .\la{(1)}\ea
The functions $\psi_{0}(x)$, $\psi_{1}(x)$ can be obtained by analytical continuation of
$\psi(x;k)$ in $k$,
\be \lim\limits_{k\to\pm
i\alpha}[(k^2+\alpha^2)\psi(x;k)]=\mp\sqrt{\alpha\over\pi}
\,\psi_{0}(x),\ee \be\lim\limits_{k\to\pm
i\alpha}\Big\{{1\over{2k}}{\partial
\over{\partial
k}}\big[(k^2+\alpha^2)\psi(x;k)\big]\Big\}=\mp\sqrt{\alpha\over\pi}
\Big[\psi_{1}(x)-{{1\mp2\alpha
z}\over{4\alpha^2}}\psi_{0}(x)\Big].\ee
For this model, the resolution of identity built of eigenfunctions
and associated functions of $h$ can be obtained by conventional
Green function methods,
\be
\delta(x-x')=\int\limits_{-\infty}^{+\infty}\psi(x;k)\psi(x';-k)\,dk+
\psi_{0}(x)\psi_{1}(x')+\psi_{1}(x)\psi_{0}(x') .\la{razl1}\ee
With the help of Dirac notations,
\be\langle x|\psi,k\rangle=\psi(x;k),\qquad\langle
x|\tilde\psi,k\rangle=
\psi^*(x;-k),\ee
\be\langle x|\psi_{0,1}\rangle=\psi_{0,1}(x),\qquad
\langle x|\hat\psi_{0,1}\rangle=
\psi_{0,1}^*(x),\ee
this resolution of identity can be represented in the operator form,
\be
I=\int\limits_{-\infty}^{+\infty}|\psi,k\rangle\langle\tilde\psi,k|\,dk+
|\psi_0\rangle\langle\hat\psi_1|+|\psi_1\rangle\langle\hat\psi_0|.\la{razl2}\quad
\ee
Evidently the basis $|\hat\psi_{1,2}\rangle$
corresponds  to the basis $|\hat\psi_n,a,i\rangle$ of Sec. 4  and therefore gives the $t$-symmetric spectral decomposition for the Hamiltonian,
\be h= \int\limits_{-\infty}^{+\infty}k^2|\psi,k\rangle\langle\tilde\psi,k|
\,dk-\alpha^2|\psi_0\rangle\langle\hat\psi_1|-
\alpha^2|\psi_1\rangle\langle\hat\psi_0|+
|\psi_0\rangle\langle\hat\psi_0| .
\ee

The diagonalization of the resolution of identity \gl{razl2} may be arranged in two different ways.
First one can exploit the scheme of Sec. 3 performing re-numeration of
certain elements of conjugated basis,
$$h^{\dagger}|\tilde\psi_1\rangle=\lambda_1|\tilde\psi_1\rangle,\qquad
(h^{\dagger}-\lambda_1)|\tilde\psi_0\rangle=|\tilde\psi_1\rangle,\qquad
h^{\dagger}|\tilde\psi,k\rangle=k^2|\tilde\psi,k\rangle, $$
namely,
\be\langle x|\psi,k\rangle=\psi(x;k),\qquad\langle
x|\tilde\psi,k\rangle=
\psi^*(x;-k),\ee
\be\langle x|\psi_{0,1}\rangle=\psi_{0,1}(x),\qquad
\langle x|\tilde\psi_{0,1}\rangle=
\psi_{1,0}^*(x).\ee
With this notation the resolution of identity reads,
\be
I=\int\limits_{-\infty}^{+\infty}|\psi,k\rangle\langle\tilde\psi,k|\,dk+
|\psi_0\rangle\langle\tilde\psi_0|+|\psi_1\rangle\langle\tilde\psi_1|.\la{razl3}\quad
\ee
We stress that $|\tilde\psi_{1,2}\rangle$ are related
to the basis $|\tilde\psi_n,a,i\rangle$  in Sec. 3.
The relevant spectral decomposition of the Hamiltonian takes the quasi-diagonal form with one Jordan cell,
\be
 h=\int\limits_{-\infty}^{+\infty}k^2|\psi,k\rangle\langle\tilde\psi,k|
\,dk-\alpha^2|\psi_0\rangle\langle\tilde\psi_0|-
\alpha^2|\psi_1\rangle\langle\tilde\psi_1|+
|\psi_0\rangle\langle\tilde\psi_1| .
\ee
On the other hand, the resolution of identity \gl{razl2} can be diagonalized by complex non-degenerate rotations , {\it i.e.} by using the construction of Sec. 4 . The relevant basis is given by,
\ba \fl &&\Psi_1(x)={1\over\sqrt2}[\varkappa\psi_{0}(x)+{{\psi_{1}(x)}\over
\varkappa}],\quad\Psi_2(x)={i\over\sqrt2}[\varkappa\psi_{0}(x)-
{{\psi_{1}(x)}\over\varkappa}],\quad\psi_{0,1}=\varkappa^{\mp1}{{\Psi_1\mp i\Psi_2}\over\sqrt2},\nonumber\\
\fl &&\int\limits_{-\infty}^{+\infty}\!\Psi_{1,2}^2(x)\,dx=1,\quad\int\limits_{-\infty}^{+\infty}\!
\Psi_1(x)\Psi_2(x)\,dx=0,\quad \int\limits_{-\infty}^{+\infty}\!\Psi_{1,2}(x)\psi(x;k)\,dx=0 ,\ea
where $\varkappa$ is an arbitrary constant .  The resolution of identity becomes diagonal ,
$$
\delta(x-x')=\int\limits_{-\infty}^{+\infty}\psi(x;k)\psi(x';-k)\,dk+
\Psi_{1}(x)\Psi_{1}(x')+\Psi_{2}(x)\Psi_{2}(x') , $$
or
in the operator form,
\be I=\int\limits_{-\infty}^{+\infty}|\psi,k\rangle\langle\tilde\psi,k|\,dk+
|\Psi_1\rangle\langle\tilde\Psi_1|+|\Psi_2\rangle\langle\tilde\Psi_2|,\label{razl4}\ee
where again the Dirac notations have been used,
\be\langle x|\Psi_{1,2}\rangle=\Psi_{1,2}(x),\qquad
\langle x|\tilde\Psi_{1,2}\rangle=\Psi^*_{1,2}(x) .\ee
Accordingly, the manifestly $t$-symmetric spectral decomposition of $h$ can be easily obtained,
\ba \fl &&h=\int\limits_{-\infty}^{+\infty}k^2|\psi,k\rangle\langle\tilde\psi,k|
\,dk\nonumber\\\fl &&-(\alpha^2-{1\over{2\varkappa^2}})|\Psi_1\rangle\langle\tilde\Psi_1|-
(\alpha^2+{1\over{2\varkappa^2}})|\Psi_2\rangle\langle\tilde\Psi_2|-
{i\over{2\varkappa^2}}|\Psi_2\rangle\langle\tilde\Psi_1|-
{i\over{2\varkappa^2}}|\Psi_1\rangle\langle\tilde\Psi_2| . \label{mansym}\ea
Notice that it cannot be diagonalized further, since the symmetric $2\times2$ matrix in \gl{mansym},
\be
\left(\begin{array}{ccc}
-\alpha^2 + \frac{1}{2\varkappa^2} & &- \frac{i}{2\varkappa^2}\\
&&\\
- \frac{i}{2\varkappa^2} & &-\alpha^2 - \frac{1}{2\varkappa^2} \end{array} \right) ,
\ee
has one degenerate eigenvalue $-\alpha^2$ and
possesses only one eigenvector ${\bf e}^t = (1, -i),\, $ with zero norm ${\bf e}^t\cdot{\bf e} = 0$.
Its existence means that the orthogonal non-degenerate matrix required for diagonalization cannot be built conventionally from a set of eigenvectors.
Evidently, this vector ${\bf e}$ maps the pair of basis functions $\Psi_1, \Psi_2$ into the self-biorthogonal eigenstate of the Hamiltonian $\psi_0$.  However its  partner in the biorthogonal basis is $\psi_1$  with $\langle \hat \psi_1| \psi_0\rangle = 1$. Thus the existence of the zero norm vector ${\bf e}$ does not entail the breakdown of resolution of identity.

\subsection{Level coalescence for complex coordinates}
The Hamiltonian $h$ with a Jordan cell for bound
state \gl{ham1} can be obtained as a limiting case,
of the Hamiltonian $h_\beta$ with two non-degenerate
bound states (of algebraic multiplicity~1), corresponding $\beta=0$,
\ba \fl && h_\beta = -\partial^2 \\ \fl &&-16\alpha^2\frac{
\frac{\alpha^2+\beta^2}{2\alpha\beta}
\sh(2\alpha x)\sh(2\beta(x-z))-2\ch^2(\alpha x)\ch(2\beta(x-z))+
2\sh^2(\beta(x-z))}
{[\sh(2\alpha x)+\frac{\alpha}{\beta}\sh(2\beta(x-z))]^2},\nonumber\ea
$$z\in\mathbb C,\qquad  {\rm{ Im}}\, z\ne0, \qquad  \alpha>0\quad ({\rm
{or}}\,\,\, -i\alpha>0),\qquad
0\le\beta<{\pi\over{2{\rm{Im}}\, z}},\qquad\beta\ne\alpha .$$
For this Hamiltonian $h_\beta$ there are two normalized eigenfunctions for bound states,
\ba &&\psi_+(x)=\sqrt{2}i\alpha\sqrt{{1\over\beta}+{1\over\alpha}}\,\cdot {{\ch\big((\alpha-\beta)x+\beta
z\big)}
\over{\sh(2\alpha x)+{\alpha\over\beta}\sh(2\beta(x-z))}},
\no
&&\psi_-(x)=\sqrt{2}\alpha\sqrt{{1\over\beta}-{1\over\alpha}}\,\cdot {{\ch\big((\alpha+\beta)x-\beta
z\big)}
\over{\sh(2\alpha x)+{\alpha\over\beta}\sh(2\beta(x-z))}},\ea
with eigenvalues,
\be h_\beta\psi_\pm=\lambda_\pm\psi_\pm,\qquad\lambda_\pm=-(\alpha\pm \beta)^2 .\ee
Eigenfunctions of $h_\beta$ for continuous spectrum take the form
\be\psi(x;k)={{[\alpha^2+\beta^2+k^2+ik{{W'(x)}\over{W(x)}}-{1\over2}{{W''(x)}
\over{W(x)}}]e^{ikx}}\over{\sqrt{2\pi}\sqrt{(k^2+\alpha^2+\beta^2)^2-4\alpha^2\beta^2}}},\la{psixk2}\ee
$$k\in\mathbb R,\qquad h_\beta\psi(x;k)=k^2\psi(x;k),\qquad W(x)=\sh(2\alpha
x)+{\alpha\over\beta}\sh(2\beta(x-z)),$$
where the branch of
$\sqrt{(k^2+\alpha^2+\beta^2)^2-4\alpha^2\beta^2}$
is defined by the condition
$$\sqrt{(k^2+\alpha^2+\beta^2)^2-4\alpha^2\beta^2}=k^2+o(k^2),\qquad
k\to\infty$$
in the complex $k$-plane with cuts, linking branch points situated in the upper
(lower) half-plane.
One can show that the biorthogonal relations hold,
$$\int\limits_{-\infty}^{+\infty}\psi^2_\pm(x)\,dx=1,\qquad
\int\limits_{-\infty}^{+\infty}\psi_+(x)\psi_-(x)\,dx=0,\qquad
\int\limits_{-\infty}^{+\infty}\psi_\pm(x)\psi(x;k)\,dx=0.$$

The analytical continuation of eigenfunctions for continuous spectrum provides the bound state
functions,
\ba \fl &&\lim\limits_{k\to\pm
i(\alpha+\beta)}[\sqrt{(k^2+\alpha^2+\beta^2)^2-4\alpha^2\beta^2}\,\psi(x;k)]=
\pm {{2i\alpha\beta}\over\sqrt\pi}\sqrt{{1\over\beta}
+{1\over\alpha}}\,e^{\mp\beta z}\psi_+(x), \no \fl &&\lim\limits_{k\to\pm
i(\alpha-\beta)}[\sqrt{(k^2+\alpha^2+\beta^2)^2-4\alpha^2\beta^2}\,\psi(x;k)]=
\mp {{2\alpha\beta}\over\sqrt\pi}\sqrt{{1\over\beta}
-{1\over\alpha}}\,e^{\pm\beta z}\psi_-(x) . \ea
Now let us coalesce two levels $\lambda_\pm$  in the limit of $\beta \rightarrow 0$ .
One can see that the eigenfunction $\psi_{0}(x)$
and associated function
$\psi_{1}(x)$ of $h$ (see Subsec. 5.1.) can be derived from
$\psi_\pm(x)$ as follows
\ba &&\psi_{0}(x)=-2i\sqrt\alpha\lim\limits_{\beta\to0}[\sqrt\beta\psi_+(x)]=
2\sqrt\alpha\lim\limits_{\beta\to0}[\sqrt\beta\psi_-(x)],\no &&\psi_{1}(x)=2\sqrt{\alpha}\lim\limits_{\beta\to0}
{{{\partial\over{\partial\beta}}\big[\sqrt\beta\big(\psi_-(x)+i\psi_+(x)\big)\big]}
\over{{\partial\over{\partial\beta}}(\lambda_--\lambda_+)}}.\ea
Resolution of identity for $\beta\ne0$ takes the conventional  form ,
$$\delta(x-x')=\psi_+(x)\psi_+(x')+\psi_-(x)\psi_-(x')+
\int\limits_{-\infty}^{+\infty}\psi(x;k)\psi(x;-k)\,dk$$
and in the limit of  $\beta \rightarrow 0$ one can reveal that in the case of $\alpha>0$
\be \lim\limits_{\beta\to0}[\psi_+(x)\psi_+(x')+\psi_-(x)\psi_-(x')]=
\psi_{0}(x)\psi_{1}(x')+\psi_{1}(x)\psi_{0}(x') ,\ee
{\it i.e.} the resolution of identity \gl{razl1} is reproduced.

\section{Puzzles with zero-binorm bound states in the continuum}
In what follows we develop another type of models
in which the continuous spectrum is
essentially involved in non-diagonal part of a Hamiltonian and
elaborate the resolution of identity. First we built the model with 
self-orthogonal bound state which however
is essentially entangled with the lower end of continuous spectrum. As a consequence, 
the self-orthogonality does not lead to infinite average values of observables 
like kinetic or potential energies if these
averages are treated with the help of wave packet regularization.

Conventionally the continuous spectrum physics deals with reflection and transmission
coefficients whose definition implies the existence of two linearly
independent scattering solutions for a given spectral parameter.
The second model provides an example when this is not realized for a
non-Hermitian Hamiltonian defined on the whole axis.

\subsection{Non-Hermitian Hamiltonian with normalizable bound state at the continuum threshold}
Let us now consider the Hamiltonian
\be h=-\partial^2+{2\over{(x-z)^2}},\qquad{\rm{Im}}\, z\ne0 .\la{ham3}\ee
The eigenfunctions of $h$ for continuous spectrum can be explicitly found,
\be
\fl \psi(x;k)={1\over\sqrt{2\pi}}\left[1-{1\over{ik(x-z)}}\right]e^{ikx},\qquad
k\in\mathbb R\backslash\{0\}, \quad h\psi(x;k)=k^2\psi(x;k).
\la{cont3}\ee
In addition, there is a normalizable eigenfunction of $h$
at the threshold of continuous spectrum,
\be
\psi_0(x)={1\over{(x-z)}}=-\sqrt{2\pi}\lim\limits_{k\to0}[ik\psi(x;k)],\qquad
h\psi_0=0. \la{lim0}\ee
Evidently the eigenfunctions of $h$ satisfy the biorthogonality relations,
\begin{eqnarray}
\int\limits_{-\infty}^{+\infty}[ik \psi(x;k)] [-ik'\psi(x;-k')]\,dx=
k^2 \delta(k-k') , \la{1}\end{eqnarray}
where the bound state wave function is included at the bottom of continuous spectrum due to \gl{lim0}.
Thus this very eigenfunction has zero binorm,
\be
\int\limits_{-\infty}^{+\infty}\psi_0^2(x)\,dx=0 ,
\ee
raising up the puzzle of "self-orthogonality" \cite{mois2}.

In order to unravel this puzzle we examine the resolution of identity made of eigenfunctions
of $h$,
\begin{equation}\delta(x-x')=\int\limits_{\cal
L}\psi(x;k)\psi(x';-k)\,dk ,
\la{2}\end{equation}
where the contour $\cal L$ must be a proper integration path in the
complex $k$ plane
which allows to regularize the singularity in \gl{cont3} for
$k = 0$, for instance, an integration path, obtained from real axis by its
displacement
near the point $k=0$ up or down.

To reach an adequate definition of resolution of identity one can instead
use the  Newton--Leibnitz
formula  and rewrite \gl{2} in the form
$$ \delta(x-x')=\Big(\int\limits_{-\infty}^{-\varepsilon}+
\int\limits_{\varepsilon}^{+\infty}\Big)\psi(x;k)\psi(x';-k)\,dk $$
\begin{equation}-
{{\psi_0(x)\psi_0(x')}\over{\pi\varepsilon}}+{{\sin\varepsilon
(x-x')}\over{\pi(x-x')}}+{{2\sin^2[{\varepsilon\over2}(x-x')]}
\over{\pi\varepsilon(x-x_0)(x'-x_0)}},\qquad\varepsilon>0 .\la{3}\end{equation}
One can show that the limit of the 3rd term of the right side of
\gl{3} (as
a distribution function) at $\varepsilon\downarrow0$ is zero for
any test
function from $C_{\mathbb R}^\infty\cap L^2({\mathbb R})$ but
the limit of the last term of the right side of \gl{3} for
$\varepsilon\downarrow0$
is zero only for test functions from $C_{\mathbb R}^\infty\cap
L^2({\mathbb R};|x|^\gamma)$, $\gamma>1$. Thus for test functions from
$C_{\mathbb R}^\infty\cap L^2({\mathbb R};|x|^\gamma)$, $\gamma>1$
resolution of identity can be reduced to,
\begin{equation}\delta(x-x')=\lim\limits_{\varepsilon\downarrow0}\bigg[
\Big(\int\limits_{-\infty}^{-\varepsilon}+
\int\limits_{\varepsilon}^{+\infty}\Big)\psi(x;k)\psi(x';-k)\,dk-
{{\psi_0(x)\psi_0(x')}\over{\pi\varepsilon}}\bigg] ,\la{4}\end{equation}
and for test functions from $C_{\mathbb R}^\infty\cap L^2({\mathbb R})$
to,
\ba \delta(x-x')&=&\lim\limits_{\varepsilon\downarrow0}\bigg\{
\Big(\int\limits_{-\infty}^{-\varepsilon}+
\int\limits_{\varepsilon}^{+\infty}\Big)\psi(x;k)\psi(x';-k)\,dk\no
&&- {1\over{\pi\varepsilon}}\Big[1-
2\sin^2\big({\varepsilon\over2}(x-x')\big)\Big]{\psi_0(x)\psi_0(x')}\bigg\}.\la{5}\ea
Decomposition \gl{4} seems to have a more natural form
than \gl{5},
but its right side obviously cannot reproduce the {\it normalizable}
eigenfunction
$$\psi_0(x)\not\in C_{\mathbb R}^\infty\cap L^2({\mathbb R};|x|^\gamma),
\qquad
\gamma>1$$
because of the orthogonality relations \gl{1}. Indeed, the identity holds,
\be \lim\limits_{\varepsilon\downarrow0}
\int\limits_{-\infty}^{+\infty}{2\over{\pi\varepsilon}}{\sin^2\big({\varepsilon\over2}(x-x')\big)}
\psi_0^2(x)\psi_0(x')\,dx=\lim\limits_{\varepsilon\downarrow0}[e^{-i\varepsilon
x'}\psi_0(x')]=\psi_0(x').\ee
Hence it is the 3rd term in the right side of \gl{5} that provides the
opportunity to reproduce $\psi_0(x)$ and thereby to complete the resolution of identity. Thus one concludes that the state $\psi_0(x)$ is inseparable from the bottom of continuous spectrum and the resolution of identity in this sense is not diagonal.

We notice that the Hamiltonian \gl{ham3} is PT-symmetric and can be derived from the Hamiltonian \gl{ham1} in the limit
$\alpha\to 0$ but the parameter $z$ must be taken as a half of $z$ from \gl{ham1}.

We also remark that the Hamiltonian \gl{ham3} makes sense also for arbitrary coupling constants of
"centrifugal" potential, and for the following set,
\be h=-\partial^2 +{{n(n+1)}\over{(x-z)^2}}\ ,\ee
with positive  $n$,  the Jordan cell, spanned by
$\big[{{n+1}\over2}\big]$ normalizable eigenfunction and associated
functions, appears at the threshold of continuous spectrum,
\ba &&h\psi_0=0 ,\qquad h\psi_j=\psi_{j-1},\qquad j=0,\ldots,
\big[{{n-1}\over2}\big],\no &&
\psi_j(x)={{(2(n-j)-1)!!}\over{(2j)!!(2n-1)!!(x-z)^{n-2j}}}.\ea
All these zero-energy bound and associated states have zero binorms and are biorthogonal to each other
(the multiple puzzle of "self-orthogonality").
Resolution of identity in such cases can be derived in a similar way although its form
will be more cumbersome.
\subsection{Expectation values (e.v.) of kinetic and potential energies in the vicinity of zero-energy bound state}
As the binorm of the bound state \gl{lim0} vanishes it seems that the quantum averages of basic observables like the kinetic $K$ or
potential $V$ energy in this system described by the Hamiltonian \gl{ham3} $h= K + V$ tend to diverge. But it is, in fact, not the case.
Indeed, the e.v.'s of these observables vanish as well,
\be
h \psi_0 (x) = 0,\quad \langle\tilde\psi_0|
V|\psi_0\rangle = -  \langle\tilde\psi_0|
K|\psi_0\rangle = \int\limits_{-\infty}^\infty dx \frac{2}{(x-z)^4} = 0 . 
\ee
Thus one comes to the classical uncertainty of $0/0$ type. In order to unravel it one has to built a wave packet which reproduces the  function $\psi_0$ in the limit of its form parameters.  We choose the Gaussian wave packet,
\ba
&&\psi_\epsilon (x) =  \int\limits_{-\infty}^\infty  \frac{dk}{\sqrt{\pi \epsilon}} \Big( - ik + \frac{1}{x-z}\Big) \exp\Big(ikx - \frac{k^2}{\epsilon}\Big) \no &&=  \Big( - \partial + \frac{1}{x-z}\Big)  \exp\Big( - \epsilon \frac{x^2}{4}\Big) =
 \Big( \epsilon\frac{x}{2} + \frac{1}{x-z}\Big)  \exp\Big( - \epsilon \frac{x^2}{4}\Big) , \la{packet}
\ea
which evidently approaches uniformly  $\psi_0$ when $\epsilon\downarrow0$ .  The binorm of this wave packet ,
\be
\langle{\tilde\psi}_\epsilon | \psi_\epsilon \rangle =  \sqrt{\frac{\pi}{8}}  \epsilon^{1/2}, \la{epsnorm}
\ee
rapidly vanishes when $\epsilon\downarrow0$ .

In turn the e.v. of the total energy,
\be
\langle{\tilde\psi}_\epsilon |H| \psi_\epsilon\rangle =  \sqrt{\frac{9\pi}{128}}  \epsilon^{3/2},
\ee
decreases with $\epsilon\downarrow0$ faster than the normalization \gl{epsnorm} . The e.v. of the potential energy,
\be
\langle{\tilde\psi}_\epsilon |V| \psi_\epsilon\rangle =  - \sqrt{\frac{25\pi}{36}}  \epsilon^{3/2},
\ee
behaves  as the total energy and therefore the e.v. for kinetic energy decreases also as $\epsilon^{3/2}$,  much faster than the binorm \gl{epsnorm} . Thus 
one concludes that  their ratios, {\it i.e.} the quantum averages of kinetic and potential energies vanish for the self-orthogonal bound state in contrast to the superficial divergence . Therefore the puzzle with self-orthogonality
is resolved.

\subsection{Hamiltonian with bound state in continuum}
Let us force the bound state energy $-\alpha^2$ in the Hamiltonian \gl{ham1} to move towards the continuous spectrum, $\alpha\to i\alpha$ . Then for the Hamiltonian,
\be
\fl h=-\partial^2+16\alpha^2{{\alpha(x-z)\sin(
2\alpha x)+2\cos^2(\alpha x)}\over{[\sin(2\alpha x)+
2\alpha(x-z)]^2}};\qquad\alpha>0,\quad z\in\mathbb C,
\quad {\rm{Im}}\,z\ne0 ,\ee
on the level $\lambda_1=\alpha^2$ in the continuous spectrum, one finds the Jordan cell, spanned by the
normalizable eigenfunction $\psi_{0}(x)$ and the associated function
$\psi_{1}(x)$, whose asymptotics for $x\to\pm\infty$ correspond
to superposition of incoming and outgoing waves (standing wave),
\be \fl \psi_{0}(x)={{\cos(\alpha x)}\over{\sin(2\alpha
x)+2\alpha(x-z)}},
\qquad\psi_{1}(x)={{{2\alpha}{{(x-z)}}\sin(\alpha x)+\cos(\alpha x)}
\over{{4\alpha^2}[\sin(2\alpha x)+2\alpha(x-z)]}},\ee
\be \fl h\psi_{0}=\lambda_1\psi_{0},\qquad(h-\lambda_1)
\psi_{1}=\psi_{0}.\ee
The asymptotics of the associated state is given by the standing wave,
$$\psi_{1}(x)={i\over{8\alpha^2}}\big[e^{-i\alpha x}-e^{i\alpha
x}\big]+O\Big({1\over x}\Big),\qquad x\to\pm\infty,$$
but it  does not appear  in the resolution of identity (see below). Thereby  this associated state does not belong to the physical state space.

In turn the eigenfunctions of $h$ for the scattering spectrum read,
\be \psi(x;k)={1\over\sqrt{2\pi}}\left[1+{{ik}\over{k^2-\alpha^2}}
{{W'(x)}\over{W(x)}}-{1\over{2(k^2-\alpha^2)}}{{W''(x)}\over{W(x)}}
\right]e^{ikx},\la{psixk4}\ee
$$k\in\mathbb R\backslash\{\alpha,-\alpha\}, \qquad
h\psi(x;k)=k^2\psi(x;k),\qquad W(x)=
\sin(2\alpha x)+2\alpha(x-z).$$
Then one can check that eigenfunctions and associated functions of $h$ obey
the relations:
\ba&&\int\limits_{-\infty}^{+\infty}\Big[\Big(k^2 - \alpha^2\Big)\psi(x;k)\Big]\Big[\Big((k')^2 - \alpha^2\Big)\psi(x;-k')\Big]\,dx= \Big(k^2 - \alpha^2\Big)^2
\delta(k-k');\no
&& \int\limits_{-\infty}^{+\infty}\psi_{0}(x)\psi_{1}(x)\,dx=0,\qquad
\int\limits_{-\infty}^{+\infty}\psi_{1}(x)\psi(x;k)\,dx=0 .\la{3''}\ea
The last equation is understood in the sense of distributions.
The limit of
$\psi(x;k)$ at $k\to\mp\alpha$ gives the elements of the Jordan cell
$\psi_{0}(x)$ and $\psi_{1}(x)$ ,
\be\lim\limits_{k\to\mp
\alpha}[(k^2-\alpha^2)\psi(x;k)]=\mp{{4i\alpha^2}\over\sqrt{2\pi}}
\psi_{0}(x),\ee
\be\lim\limits_{k\to\mp \alpha}\Big[{1\over{2k}}{\partial\over{\partial
k}}
\big((k^2-\alpha^2)\psi(x;k)\big)\Big]=\mp{{4i\alpha^2}\over\sqrt{2\pi}}\Big[
\psi_{1}(x)+{{1\mp2i\alpha z}\over{4\alpha^2}}\psi_{0}(x)\Big].\ee
For this model, resolution of identity made of  eigenfunctions
and associated functions of $h$ can be obtained by conventional
Green function methods,
\be \delta(x-x')=\int\limits_{\cal L}\psi(x;k)\psi(x';-k)\,dk,\ee
where $\cal L$ is an integration path in the complex momentum plane, obtained from real axis by its
simultaneous displacement
near the points $k=\pm\alpha$ up or down.

For test functions from $C_{\mathbb R}^\infty\cap
L^2({\mathbb R};|x|^\gamma)$, $\gamma>1$ this resolution of identity can be
presented in
the form,
\be
\fl \delta(x-x')=\lim_{\varepsilon\downarrow0}
\Big[\Big(\int\limits_{-\infty}^{-\alpha-\varepsilon}+
\int\limits_{-\alpha+\varepsilon}^{\alpha-\varepsilon}+
\int\limits_{\alpha+\varepsilon}^{+\infty}\Big)\psi(x;k)\psi(x';-k)\,dk
-{1\over{\pi\varepsilon\alpha}}\psi_{0}(x)\psi_{0}(x')\Big]\la{2''}\ee
and for test functions from $C_{\mathbb R}^\infty\cap L^2({\mathbb R})$ it must be extended,
$$\delta(x-x')=\lim_{\varepsilon\downarrow0}
\bigg\{\Big(\int\limits_{-\infty}^{-\alpha-\varepsilon}+
\int\limits_{-\alpha+\varepsilon}^{\alpha-\varepsilon}+
\int\limits_{\alpha+\varepsilon}^{+\infty}\Big)\psi(x;k)\psi(x';-k)\,dk$$
\be-{1\over{\pi\varepsilon\alpha}}\Big[1-
2\sin^2\big({\varepsilon\over2}(x-x')\big)\Big]\psi_{0}(x)\psi_{0}(x')\bigg\}\la{2+}\ee
(cf. with \gl{5}). One can see that the operator \gl{2''} projects away
the {\it normalizable} eigenfunction
$$\psi_{0}(x)\not\in C_{\mathbb R}^\infty\cap L^2({\mathbb R};|x|^\gamma),
\qquad
\gamma>1$$
because of the orthogonality relations \gl{3''}.
Meanwhile the operator \gl{2+} is complete and acts on this eigenfunction as identity. Thus one concludes again that the state $\psi_0(x)$ is inseparable from the continuous spectrum and the resolution of identity in this sense is not diagonal.

\section{Resolvents and scattering characteristics}
The peculiar spectral properties and the specific pattern of level degeneracy
for non-diagonalizable Hamiltonians has interesting consequences for the structure of their resolvents and scattering matrices.

In all the above examples the Green functions  can be calculated conventionally, as follows,
\be\fl G(x,x';\lambda)={{\pi i}\over\sqrt{\lambda}}
\psi(x_>;\sqrt{\lambda})\psi(x_<;-\sqrt{\lambda}),\quad
x_>=\max\{x,x'\},\quad x_<=\min\{x,x'\},\ee
where the solutions $\psi$ are made by analytical continuation of $\psi(x;k)$
in $k$ into complex plane, and
the branch of $\sqrt{\lambda}$ is uniquely defined by the
condition ${\rm{Im}}\sqrt{\lambda}\ge0$ in the plane with the cut on
positive part of
real axis.  In virtue of \gl{psixk2}
the Green function for the diagonalizable Hamiltonian of Subsec. 5.2.  has two poles of
the first order (if $\beta\ne0$) at the points
$\lambda=-(\alpha\pm\beta)^2$ where $\alpha$ can be either real or imaginary. If $\beta \rightarrow 0$ and level confluence emerges, two poles coalesce into
one pole of the second order in both cases \gl{psixk1} and \gl{psixk4} when $\lambda=\mp |\alpha|^2$ . However the
examples of Subsec. 5.1 and 6.2 have different meaning: in the first case the double pole does not appear on the physical cut $\lambda >0$ and its order
enumerates the rank of the Jordan cell. On the contrary, in the second case the double pole is placed exactly on the cut $\lambda >0$ and strictly speaking signifies the spectral pathology as it generates only one eigenstate with the eigenvalue $\lambda=|\alpha|^2$ in resolution of identity. The second state -- the associated function, represents a standing wave and does not influence the spectral decomposition.  In the example of Subsec. 6.1. the Green function has only a branch point at $\lambda=0$ of the following type $\lambda^{-3/2}$.
This  branch point can be thought of as a  confluence of the double pole in the variable  $\sqrt{\lambda}$ and the branch point of order $\lambda^{-1/2}$ .

From the explicit form of wave functions one can see that all potentials in Sec.5 and 6 are transparent as the reflection coefficient
is zero. The transmission coefficient in the non-degenerate case of Subsec. 5.2. takes the form,
\be
T(k)=\left\{\begin{array}{ccc}\frac{\beta^2+(k+i\alpha)^2}{\beta^2+(k-i\alpha)^2},
& &0<\beta<\alpha,\\
&&\\
\frac{\alpha^2+(k+i\beta)^2}{\alpha^2+(k-i\beta)^2},&&
-i\alpha>0 .\end{array}\right.
\ee
In different limits one can derive the transmission coefficients for three other Hamiltonians. Namely, when  $\beta \rightarrow 0$ and $\alpha > 0$ (the case 5.1.) the scattering is described by
\be
T(k)=\Big({{k+i\alpha}\over{k-i\alpha}}\Big)^2,
\ee
and for $\beta \rightarrow 0$ and $\alpha \rightarrow 0$ or imaginary (the cases 6.1. and 6.2.) the scattering is absent,
\be
T(k)=1 .\ee
Thus the colliding particle in such cases is not "influenced" by the bound
states in the continuum.

\section{Conclusions}
In this paper we have presented a thorough analysis of the phenomenon of apparent self-orthogonality of some eigenstates for non-Hermitian Hamiltonians. For the discrete part of energy spectrum, it has been shown that such a phenomenon should take place only for non-diagonalizable Hamiltonians the spectrum of which consists  not only of eigenfunctions and but also of associated functions. However the genuine {\it diagonal} biorthogonal basis related to the spectral decomposition of such Hamiltonians normally does not contain pairs made of the same eigenfunctions or the associated functions of the same order.  Rather they are complementary: for instance,
eigenfunctions in the direct basis are paired to those associated functions in the conjugated basis, which have the maximal order in the same Jordan cell.
One possible exception exists for Jordan cells of odd order where one basis pair consists of the same function which is {\it not} self-orthogonal.

The situation in the continuous spectrum is more subtle: namely, the spectral decomposition does not include any obvious Jordan cells and associated functions. However we have established that when a zero-binorm normalizable state
arises it remains inseparable from the nearest scattering states of the continuum and eventually the existence of this state does not destroy the completeness of
resolution of identity.

Finally, let us outline the measurability of quantum observables and, for this purpose, prepare a wave packet,
\be |\psi\rangle=\sum\!\!\!\!\!\!\!\!\int\limits_r C_r|\psi_r\rangle,\qquad
C_r={\rm{Const}}, \label{wavep}\ee
where the possibility to have continuous spectrum is made explicit in the notation
and, for brevity, all indices enumerating eigenvalues, Jordan cells and their elements are encoded in the index  $r \equiv \{n,a,j\}$.

In order to perform the quantum averaging of an operator of observable $O$ one can use the conventional Hilbert space scalar product and the complex conjugated wave function, $\langle\psi|x\rangle = \langle x|\psi\rangle^*$ . In this way one defines the wave packet of the conjugated state and, respectively, average values of the operator $O$ ,
\begin{equation} \langle\psi|=\sum\!\!\!\!\!\!\!\!\int\limits_r C_r^*\langle\psi_r|,\qquad \overline{O} ={{\langle\psi|
O|\psi\rangle}\over{\langle\psi|\psi\rangle}},\la{aver1}\end{equation}
so that average values $\overline{O}$ remain finite as $\langle\psi|\psi\rangle>0$ .

On the other hand,  the use of {\it complete} biorthogonal bases  $\{|\psi_r\rangle, \langle\tilde\psi_r|\} $ from Sec.3 (or $\{|\psi_r\rangle, \langle\hat\psi_r|\} $ from Sec.4) seems to be more suitable to describe a non-Hermitian evolution. Accordingly, to supply the wave packet binorm with a probabilistic meaning  one may introduce \cite{curt} the wave packet partner in respect to a binorm with complex conjugated coefficients, in order that its binorm were always positive,
\be \langle\tilde\psi|\equiv \sum\!\!\!\!\!\!\!\!\int\limits_r C_r^*\langle\tilde\psi_r|,\qquad\langle\tilde\psi|\psi\rangle=\sum\!\!\!\!\!\!\!\!\int\limits_r C_r^*C_r>0 .\ee
For such a definition the averages of an  operator of observable,
$\widetilde{\overline{O}} ={\langle\tilde\psi|
O|\psi\rangle}/{\langle\tilde\psi|\psi\rangle} $ cannot be infinite
and the phenomena of (pseudo) phase
transitions at the level crossing  \cite{mois2} cannot appear. In this relation the basis from Sec. 4 with $\langle\hat\psi_r|$ may be used equally well.

However one has to keep in mind that such a definition of probabilities, to some extent, depends on a particular set of biorthogonal bases. We hope to examine this approach and its applications elsewhere.

\ack
The work of A.A. and A.S. was supported by Grant RFBR 06-01-00186-a and  by the Programs RNP 2.1.1.1112 and RSS-5538.2006.2.  A.Sokolov was partially supported by
Grant INFN-IS/BO31. We acknowledge Prof. S.L. Yakovlev for useful remarks and relevant references. We are grateful to the organizers of 3rd International Workshop on
Pseudo-Hermitian Hamiltonians in Quantum Physics (June 2005, Istanbul, Turkey), especially to Prof. A.~Mostafazadeh and Prof. M. Znojil,
for financial support and opportunity to present the preliminary results of this paper.
\section*{Appendix}

Let $CL_\gamma=C^\infty_{\mathbb R}\cap L_2(\mathbb R; |x|^\gamma)$, $\gamma\geqslant 0$
be the space of test functions. The sequence $\varphi_n(x)\in CL_\gamma$ is
called convergent in $CL_\gamma$ to $\varphi(x)\in CL_\gamma$,
\be \mathop{\lim\nolimits\nolimits_\gamma}\limits_{n\to+\infty}\varphi_n(x)=\varphi(x),\ee
if
\be \lim_{n\to+\infty}\int\limits_{-\infty}^{+\infty}|\varphi_n(x)-
\varphi(x)|^2|x|^\gamma dx=0 ,\ee
and for any $x_1,x_2\in\mathbb R$, $x_1<x_2$,
\be\lim\limits_{n\to+\infty}\max\limits_{[x_1,x_2]}|\varphi_n(x)-
\varphi(x)|=0 .\ee

We shall denote the value of a functional $f$ on $\varphi\in CL_\gamma$ 
conventionally as  $(f,\varphi)$. A functional $f$ is called continuous, if for
any sequence $\varphi_n\in CL_\gamma$ convergent in $CL_\gamma$ to zero
the equality ,
\be \lim\limits_{n\to+\infty}(f,\varphi_n)=0, \ee
is valid. The space of distributions over $CL_\gamma$, {\it i.e.}
of linear continuous functionals over $CL_\gamma$ is denoted as $CL'_\gamma$.
The sequence $f_n\in CL'_\gamma$ is called convergent in $CL'_\gamma$ 
to $f\in CL_\gamma$,
\be \mathop{\lim\nolimits'_\gamma}\limits_{n\to+\infty} f_n=f,\ee
if for any $\varphi\in CL_\gamma$ the relation takes place
\be \lim\limits_{n\to+\infty}(f_n,\varphi)=(f,\varphi) .\ee

A functional $f\in CL'_\gamma$ is called regular, if there is $f(x)\in
L_2(\mathbb R;(1+|x|^\gamma)^{-1})$ such that for any $\varphi\in CL_\gamma$
the equality
\be (f,\varphi)=\int\limits_{-\infty}^{+\infty}f(x)\varphi(x)\,dx\ee
holds. In this case  we shall identify the distribution $f\in 
CL'_\gamma$ with the function $f(x)\in L_2(\mathbb R; (1+|x|^\gamma)^{-1})$.
In virtue of the Bunyakovskii inequality,
\be \Big|\int\limits_{-\infty}^{+\infty}f(x)\varphi(x)\,dx\Big|^2\leqslant
\int\limits_{-\infty}^{+\infty}{{|f^2(x)|\,dx}\over{1+|x|^\gamma}}
\int\limits_{-\infty}^{+\infty}|\varphi^2(x)|(1+|x|^\gamma)\,dx\ee
it is evident that $L_2(\mathbb R;(1+|x|^\gamma)^{-1})\subset CL'_\gamma$.

Let us notice also  that the  Dirac delta-function $\delta(x-x')$ 
belongs to $CL'_\gamma$, $\gamma\geqslant 0$.

{\bf Lemma 1.} {\it Suppose that: 1) $$\psi(x;k)={1\over\sqrt{2\pi}}
[1-{1\over{ik(x-z)}}]e^{ikx} ;$$ 2) ${\cal L}(A)$ is a path in the complex plane
of $k$, made of the segment $[-A,A]$ of real axis by deformation of its central 
part up or down of zero and the positive direction of ${\cal L}(A)$ is specified
from $-A$ to $A$; 3) $x'\in \mathbb R$, $\gamma\geqslant0$, 
$\varepsilon>0$.
Then the following relations hold,
\be \mathop{\lim\nolimits'_\gamma}\limits_{A\to+\infty}\, \int_{{\cal L}(A)}\psi(x;k)\psi(x';-k)\,dk
=\delta(x-x'),\la{ap1}\ee
\be 
\fl \mathop{\lim\nolimits'_\gamma}\limits_{A\to+\infty}\, (\int\limits_{-A}^{-\varepsilon}+
\int\limits_{\varepsilon}^A)\psi(x;k)\psi(x';-k)\,dk
=\delta(x-x')-\int_{{\cal L}(\varepsilon)}\psi(x;k)\psi(x';-k)\,dk .\la{ap2}\ee
}

{\bf Proof.}
In accordance with the Newton--Leibnitz formula one obtains,
\be\int_{{\cal L}(A)}\psi(x;k)\psi(x';-k)\,dk={1\over\pi}\Big[
{{\sin A(x-x')}\over{x-x'}}-{{\cos A(x-x')}\over{A(x-z)(x'-z)}}\Big]  .\la{ap10}\ee
The integral \gl{ap10} as  a function of $x$ belongs to
$L_2(\mathbb R;(1+|x|^\gamma)^{-1})$ and therefore to $CL'_\gamma$. Thus
to prove \gl{ap1} it is sufficient to establish that for any $\varphi(x)\in 
CL_\gamma$ the equality takes place,
\be
\fl\lim\limits_{A\to+\infty}{1\over\pi}\int\limits_{-\infty}^{+\infty}\Big[
{{\sin A(x-x')}\over{x-x'}}-{{\cos A(x-x')}\over{A(x-z)(x'-z)}}\Big]
\varphi(x)\,dx=\varphi(x') .\la{ap11}\ee
By virtue of the Bunyakovskii inequality
\be
\fl \Big|\int\limits_{-\infty}^{+\infty}{{\cos A(x-x')}\over{A(x-z)(x'-z)}}
\varphi(x)\,dx\Big|^2\leqslant{1\over A^2}\int\limits_{-\infty}^{+\infty}{{dx}
\over{|x-z|^2|x'-z|^2}}\int\limits_{-\infty}^{+\infty}|\varphi^2(x)|\,dx ,\ee
wherefrom it follows that
\ba
&& \lim\limits_{A\to+\infty}{1\over\pi}\int\limits_{-\infty}^{+\infty}\Big[
{{\sin A(x-x')}\over{x-x'}}-{{\cos A(x-x')}\over{A(x-z)(x'-z)}}\Big]
\varphi(x)\,dx\nonumber\\
&&=\lim\limits_{A\to+\infty}{1\over\pi}
\int\limits_{-\infty}^{+\infty}{{\sin A(x-x')}\over{x-x'}}
\varphi(x)\,dx.\la{ap12}\ea
In virtue  of the Riemann theorem and due to the evident inclusions $${{\varphi(x)}\over{
x-x'}}\in L_1(\mathbb R\setminus]x'-\delta,x'+\delta[), \quad {{\varphi(x)-\varphi(x')}\over{
x-x'}}\in L_1([x'-\delta,x'+\delta])$$ for any $\delta>0$ the following relations are valid, 
\ba &&\lim\limits_{A\to+\infty}
\Big(\int\limits_{-\infty}^{x'-\delta}+\int\limits_{x'+\delta}^{+\infty}\Big)
\sin A(x-x'){{\varphi(x)}\over{x-x'}}\,dx =0,\nonumber\\
&&\lim\limits_{A\to+\infty}
\int\limits_{x'-\delta}^{x'+\delta}\sin A(x-x'){{\varphi(x)-\varphi(x')}
\over{x-x'}}\,dx=0.\ea
Hence,
$$\lim\limits_{A\to+\infty}{1\over\pi}
\int\limits_{-\infty}^{+\infty}{{\sin A(x-x')}\over{x-x'}}
\varphi(x)\,dx={\varphi(x')\over\pi}\lim\limits_{A\to+\infty}
\int\limits_{x'-\delta}^{x'+\delta}{{\sin A(x-x')}\over{x-x'}}\,dx =\varphi(x')$$
and in view of \gl{ap12}  Eqs. \gl{ap11} and  \gl{ap1} hold.
Eq. \gl{ap2} follows from Eq.\gl{ap1} and from additivity of classical
path integrals. Lemma 1 is proved.

{\bf Lemma 2.} {\it For any $x'\in\mathbb R$ and $\gamma\geqslant 0$ the relation,
\be\mathop{\lim\nolimits'_\gamma}\limits_{\varepsilon\downarrow0}{{\sin\varepsilon(x-x')}
\over{x-x'}}=0,\qquad\gamma\geqslant 0,\la{ap3}\ee
takes place .}

{\bf Proof.} It is true that $${{\sin\varepsilon(x-x')}\over{x-x'}}
\in L_2(\mathbb R;(1+|x|^\gamma)^{-1})\subset CL'_\gamma, \quad \gamma\geqslant0 .$$
Thus, to prove the lemma it is sufficient to establish that for any $\varphi(x)\in 
CL_\gamma$, $\gamma\geqslant0$ the relation
\be\lim\limits_{\varepsilon\downarrow0}\int\limits_{-\infty}^{+\infty}
{{\sin\varepsilon(x-x')}\over{x-x'}}\varphi(x)\,dx=0\ee
is valid. But its validity follows from the Bunyakovskii
inequality
$$\Big|\int\limits_{-\infty}^{+\infty}{{\sin\varepsilon(x-x')}
\over{x-x'}}\varphi(x)\,dx\Big|^2\leqslant\int\limits_{-\infty}^{+\infty}
{{\sin^2\varepsilon(x-x')}\over{(x-x')^2}}\,dx
\int\limits_{-\infty}^{+\infty}|\varphi^2(x)|\,dx=$$
\be\varepsilon\int\limits_{-\infty}^{+\infty}
{{\sin^2\tau}\over\tau^2}\,d\tau
\int\limits_{-\infty}^{+\infty}|\varphi^2(x)|\,dx\to0,\qquad
\varepsilon\downarrow0.\ee
Lemma 2 is proved.

{\bf Lemma 3.} {\it For any $z\in\mathbb C$, ${\rm{Im}}\,z\ne0$, $x'\in\mathbb R$ and 
$\gamma > 1$ the relation holds,
\be\mathop{\lim\nolimits'_\gamma}\limits_{\varepsilon\downarrow0}{{\sin^2[{\varepsilon\over2}
(x-x')]}\over{\varepsilon(x-z)(x'-z)}}=0,\qquad\gamma>1.\la{ap4}\ee
}

{\bf Proof.} It is evident that $${{\sin^2[{\varepsilon\over2}
(x-x')]}\over{\varepsilon(x-z)(x'-z)}}\in L_2(\mathbb R;(1+|x|^\gamma)^{-1})
\subset CL'_\gamma,\quad \gamma>1 .$$ Thus, to prove the lemma it is sufficient
to establish that for any $\varphi(x)\in CL_\gamma$, $\gamma>1$ the equality
\be\lim\limits_{\varepsilon\downarrow0}\int\limits_{-\infty}^{+\infty}
{{\sin^2[{\varepsilon\over2}(x-x')]}\over{\varepsilon(x-z)(x'-z)}}
\varphi(x)\,dx=0\ee
holds. This equality can be obtained from the chain of inequalities
$$\Big|\int\limits_{-\infty}^{+\infty}{{\sin^2[{\varepsilon\over2}(x-x')]}
\over{\varepsilon(x-z)(x'-z)}}\varphi(x)\,dx\Big|^2\leqslant$$
$$\int\limits_{-\infty}^{+\infty}
{{\sin^4[{\varepsilon\over2}(x-x')]\,dx}\over{\varepsilon^2|x-z|^2|x'-z|^2
(1+|x|^\gamma)}}
\int\limits_{-\infty}^{+\infty}|\varphi^2(x)|(1+|x|^\gamma)\,dx\leqslant$$
\be
\fl \int\limits_{-\infty}^{+\infty}
{{(\varepsilon/2)^{2+\min\{2,(\gamma-1)/2\}}|x-x'|^{2+{\min\{2,(\gamma-1)/2\}}}dx}
\over{\varepsilon^2|x-z|^2|x'-z|^2(1+|x|^\gamma)}}
\int\limits_{-\infty}^{+\infty}|\varphi^2(x)|(1+|x|^\gamma)\,dx\to0,\quad
\varepsilon\downarrow0, \ee
derived with the help of Bunyakovskii inequality and trivial 
inequalities $|\sin\tau|\leqslant1$, $|\sin\tau|\leqslant|\tau|$, $\tau\in\mathbb 
R$. Lemma 3 is proved.

{\bf Corollary 1.}
Let's define
\be \int_{\cal L}\psi(x;k)\psi(x';-k)\,dk=
 \mathop{\lim\nolimits'_\gamma}\limits_{A\to+\infty}\int_{{\cal L}(A)}\psi(x;k)\psi(x';-k)\,dk,
\la{ap5}\ee
where $\cal L$ is a path, made by deformation of real axis near zero up or down.
Then in view of \gl{ap1} the resolution of identity \gl{2} holds.

{\bf Corollary 2.} Using the Newton--Leibnitz formula, one can rewrite integral
$\int_{{\cal L}(\varepsilon)}\psi(x;k)\psi(x';-k)\,dk$ in the form
\be 
\fl \int_{{\cal L}(\varepsilon)}\psi(x;k)\psi(x';-k)\,dk= -
{1\over{\pi\varepsilon(x-z)(x'-z)}}+{{\sin\varepsilon(x-x')}\over{\pi(x-x')}}+
{{2\sin^2[{\varepsilon\over2}(x-x')]}\over{\pi\varepsilon(x-z)(x'-z)}}.
\la{ap6}\ee
Thus, if by definition
\be 
\fl (\int\limits_{-\infty}^{-\varepsilon}+
\int\limits_{\varepsilon}^{+\infty})\psi(x;k)\psi(x';-k)\,dk=
 \mathop{\lim\nolimits'_\gamma}\limits_{A\to+\infty}(\int\limits_{-A}^{-\varepsilon}+
\int\limits_{\varepsilon}^A)\psi(x;k)\psi(x';-k)\,dk,\la{ap7}\ee
then due to Eqs. \gl{ap2}, \gl{ap3}, \gl{ap4} and \gl{ap6} the resolutions
of identity \gl{4} and \gl{5} are valid.

\end{document}